\documentclass[prl, twocolumn,10pt,longbibliography,superscriptaddress]{revtex4-2}
\usepackage{textcomp}
\usepackage{amsmath}
\usepackage{amsfonts}
\usepackage{calc}
\usepackage{mathrsfs}
\usepackage{amssymb}
\usepackage{amsmath}
\usepackage{array}
\usepackage{color}
\usepackage{bm}
\usepackage{graphicx}
\usepackage{hyperref}
\usepackage{booktabs}
\usepackage[scr=boondoxo,scrscaled=1.05]{mathalfa}
\usepackage[dvipsnames]{xcolor}

\newcommand{\e}{\epsilon}

\newcommand{\F}{\mathcal{F}}
\newcommand{\nF}{\hat{\mathcal{F}}}

\begin{document}
\title{Gravitational-wave energy flux for compact binaries through\\second order in the mass ratio}
\author{Niels Warburton}
\affiliation{School of Mathematics and Statistics, University College Dublin, Belfield, Dublin 4, Ireland, D04 V1W8}
\author{Adam Pound} 
\affiliation{School of Mathematical Sciences and STAG Research Centre, University of Southampton, Southampton, United Kingdom, SO17 1BJ}
\author{Barry Wardell}
\affiliation{School of Mathematics and Statistics, University College Dublin, Belfield, Dublin 4, Ireland, D04 V1W8}
\author{Jeremy Miller}
\affiliation{Department of Physics, Ariel University, Ariel 40700, Israel}
\author{Leanne Durkan}
\affiliation{School of Mathematics and Statistics, University College Dublin, Belfield, Dublin 4, Ireland, D04 V1W8}
\date{\today}

\providecommand{\NW}[1]{{\textcolor{Red}{\texttt{NW: #1}}}}
\providecommand{\AP}[1]{{\textcolor{Red}{\texttt{AP: #1}}}}
\providecommand{\LD}[1]{{\textcolor{Red}{\texttt{LD: #1}}}}
\providecommand{\BW}[1]{{\textcolor{Red}{\texttt{BW: #1}}}}

\renewcommand{\e}{\epsilon}

\begin{abstract}
Within the framework of self-force theory, we compute the gravitational-wave energy flux through second order in the mass ratio for compact binaries in quasicircular orbits.
Our results are consistent with post-Newtonian calculations in the weak field and they agree remarkably well with numerical-relativity simulations of comparable-mass binaries in the strong field.
We also find good agreement for binaries with a spinning secondary or a slowly spinning primary.
Our results are key for accurately modelling extreme-mass-ratio inspirals and will be useful in modelling intermediate-mass-ratio systems. 
\end{abstract}

\maketitle

\textit{Introduction.}
Advances in gravitational wave (GW) astronomy will come from the development of experimental apparatus, data analysis algorithms, and theoretical waveform templates.
For the inspiral and merger of compact binaries, the latter necessitates solving the two-body problem in general relativity.
Over the decades, various approaches have been developed to do so by obtaining approximate solutions to the Einstein field equations.
Post-Newtonian (PN) theory applies in the weak field, making it valid early in the inspiral, when the objects are far apart  \cite{Blanchet:14}.
Effective-one-body theory extends PN theory's domain of validity and allows for calibration with strong-field data in the late inspiral, close to merger \cite{Buonanno-Damour:99}.
In the strong field no analytic approximations suffice and usually one must turn to numerical relativity (NR) simulations~\cite{Pretorius:2005gq,Campanelli:2005dd}.
Though these provide an exact result (modulo numerical error), their high computational burden means they are restricted to near-comparable-mass binaries and a few tens to hundreds of GW cycles.

When the ratio of the mass of the smaller (secondary) object to that of the primary is small, it is natural to turn to the gravitational self-force (GSF) approach and black hole perturbation theory (BHPT) \cite{Barack:2018yvs,Pound:2021qin}. 
Within this method the binary's spacetime metric is expanded in powers of the (small) mass ratio around that of the primary, larger object.
Traditionally, the GSF approach has been used to model extreme-mass-ratio inspirals (EMRIs): binaries where a compact object inspirals into a supermassive black hole with a mass ratio of $1:10^5$ or smaller.
These systems are key sources for the future Laser Interferometer Space Antenna, LISA \cite{2017arXiv170200786A}.

In order to extract EMRI signals from the LISA data stream, and to enable precision tests of general relativity~\cite{Gair:2012nm}, GSF calculations must be carried through to second order in the mass ratio~\cite{Hinderer-Flanagan:08}.
The calculation of first-order GW fluxes has been possible since the 1970s~\cite{Teukolsky:73} and has enabled the computation of adiabatic inspirals.
Within the last two decades, post-adiabatic corrections have been formulated and computed. 
These include first-order conservative corrections to the dynamics~\cite{Barack-Sago:11,vandeMeent:17b,Barack:2018yvs}, formulations at second order~\cite{Pound:12a,Gralla:12,Pound:12b,Pound:17}, and a lone calculation of a second-order quantity~\cite{Pound:2019lzj}.

\begin{figure}
	\includegraphics[width=8.5cm]{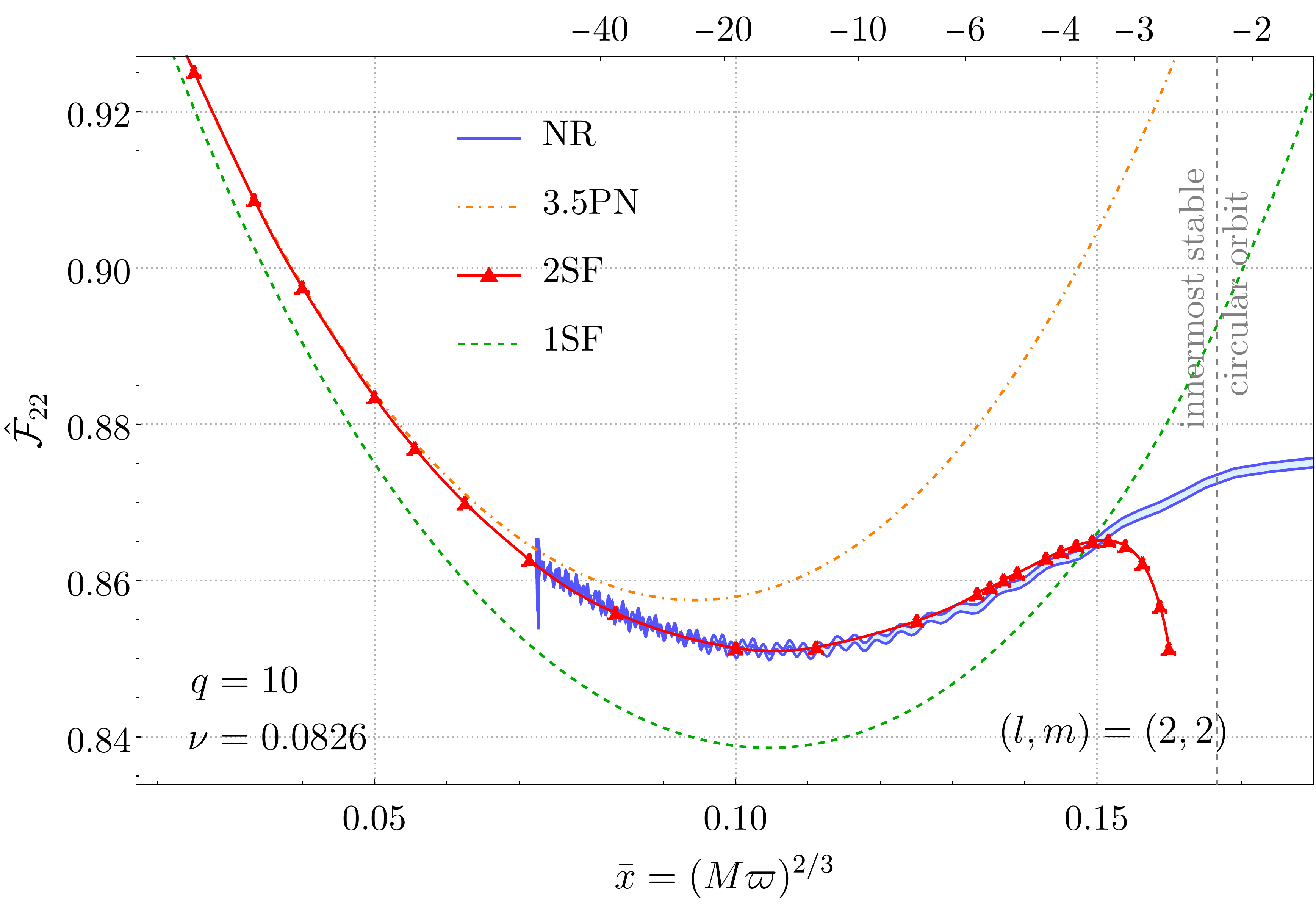}
	\caption{The gravitational-wave flux (normalized by its leading Newtonian behaviour) for a nonspinning binary as a function of the inverse orbital separation. Shown is the $(l,m)=(2,2)$ mode for a mass ratio 10:1 binary computed using the PN, NR and GSF approaches. The solid, oscillating (blue) curve shows the NR flux computed from SXS:BBH:1107 \cite{Boyle:2019kee}. The numbers along the top axis count the cycles before the peak amplitude in the NR waveform. The solid (red) curve shows the result from our second-order GSF (2SF) calculation. This agrees remarkably well with the NR result until very close to merger, where the GSF contributions diverge as the two-timescale approximation breaks down. In the weak field the second-order self-force data agrees with the 3.5PN series \cite{Faye:2012we}, shown by the (orange) dash-dotted curve. We also show the first-order self-force (1SF) result with the (green) dashed curve. The vertical, dashed (gray) line marks the location of the (geodesic) innermost stable circular orbit.
}\label{fig:GSFvsNRvsPN_l2m2}
\end{figure}

In this Letter we report the first calculation of a key physical observable that characterises a binary's post-adiabatic evolution: the flux of energy in GWs radiated to future null infinity (hereafter referred to as ``the flux'') including all contributions through second order in the mass ratio (2SF).
We focus on nonspinning binaries, but also present results for binaries where the components are spinning with a small angular momentum.

We find that the 2SF flux agrees remarkably well with NR simulations for near-comparable-mass binaries. 
This agreement holds until a few cycles from merger, when the slow-inspiral assumption in our calculation breaks down.
Figure \ref{fig:GSFvsNRvsPN_l2m2} summarizes the results of these comparisons.
It is not completely unexpected that BHPT can be pushed beyond its traditional domain of validity.
For years there has been mounting evidence that this is the case both in the conservative sector \cite{LeTiec-etal:11, LeTiec-etal:13, LeTiec:14} and via comparisons between NR and first-order GSF waveforms \cite{Rifat:2019ltp,vandeMeent:2020xgc}. Our work is the first time the 2SF flux has been computed. By comparison with NR, it strongly suggests that GSF results can be used to model intermediate-mass-ratio inspirals (IMRIs), as well as EMRIs.

We use geometrized units with $G=c=1$.
We denote the masses of the binary components by $m_1$ and $m_2$ with $m_1 \ge m_2$. 
We also define the small mass ratio $\e = m_2/m_1$, large mass ratio $q=1/\epsilon$, and symmetric mass ratio $\nu = m_1 m_2/M^2$, where $M=m_1 + m_2$.
For (anti-)aligned spinning binaries we define the dimensionless spin variables $\chi_i = S_i/m_i^2$, where $i=\{1,2\}$ and the $S_i$'s are the components of the dimensionful spin vectors in the direction of the orbital angular momentum.

\textit{Second-order self-force calculation.}
Our calculation implements the two-timescale formalism of Ref.~\cite{Miller:2020bft}. Restricting our attention to quasicircular orbits with orbital frequency $\Omega = d\phi_p/dt$, where $\phi_p(t)$ is the azimuthal angle of the orbiting secondary, we write the binary's metric as
\begin{equation}
\label{eq:pert}
 g_{\alpha\beta} +\! \sum_{m=-\infty}^\infty\! \left[\e h^{1,m}_{\alpha\beta}(\Omega) + \e^2 h^{2,m}_{\alpha\beta}(\Omega)\right]e^{-im\phi_p} +\mathcal{O}(\e^3),
\end{equation}
where $g_{\alpha\beta}$ is the Schwarzschild metric of the primary. 
The orbital frequency and the metric perturbation amplitudes $h^{n,m}_{\alpha\beta}(\Omega)$ evolve slowly, on the radiation-reaction timescale, according to Eq.~(A4) in Ref.~\cite{Miller:2020bft}, whereas the phase $\phi_p$ evolves rapidly, on the orbital timescale.
These two timescales are disparate during the inspiral, only becoming commensurate close to the innermost stable circular orbit (ISCO), where the expansion breaks down.

In order to compute the amplitudes $h^{n,m}_{\alpha\beta}$, we substitute Eq.~\eqref{eq:pert} into the Einstein equation and solve order-by-order in $\e$. The secondary is incorporated using an analytically known puncture, which diverges on the secondary's trajectory but captures the dominant part of the physical, finite metric in the secondary's local neighborhood~\cite{Pound:12b,Pound-Miller:14} (equivalent to treating it as a point mass~\cite{Upton:2021oxf}).
Working in the Lorenz gauge and decomposing $h^{n,m}_{\alpha\beta}$ onto a basis of tensor spherical harmonics with modal indices $lm$, as in Eq.~(A3) of~\cite{Miller:2020bft},
we reduce the field equations to a set of ordinary differential equations  for the radial coefficients, given explicitly in Eqs.~(152)--(153) of Ref.~\cite{Miller:2020bft}, as well as obtaining evolution equations for the mass and spin of the primary, Eqs.~(227) and (242) of~\cite{Miller:2020bft}. These equations can be solved for the $lm$ modes of $h^{n,m}_{\alpha\beta}$ at any values of $\Omega$ without knowledge of $\phi_p$, with the system's slow evolution accounted for through source terms proportional to $d\Omega/dt$ in the second-order field equations.

We compute the source in the second-order field equations and derive boundary conditions using the techniques developed in Refs.~\cite{Miller-Wardell-Pound:16,Pound:15c,Spiers_in_prep}.
Key inputs for the source are $h^{1,m}_{\alpha\beta}$, $\partial_{\Omega} h^{1,m}_{\alpha\beta}$, and the first-order GSF, all of which we compute numerically~\cite{Akcay-Warburton-Barack:13,Miller_in_prep,Durkan_in_prep}.
With these in hand, we numerically solve the radial field equations for each $lm$ mode of $h^{2,m}_{\alpha\beta}$ on hyperboloidal slices~\cite{Miller:2020bft} using the method of variation of parameters~\cite{Wardell-Warburton:15,Miller_in_prep}.

\begin{figure}
	\includegraphics[width=8.5cm]{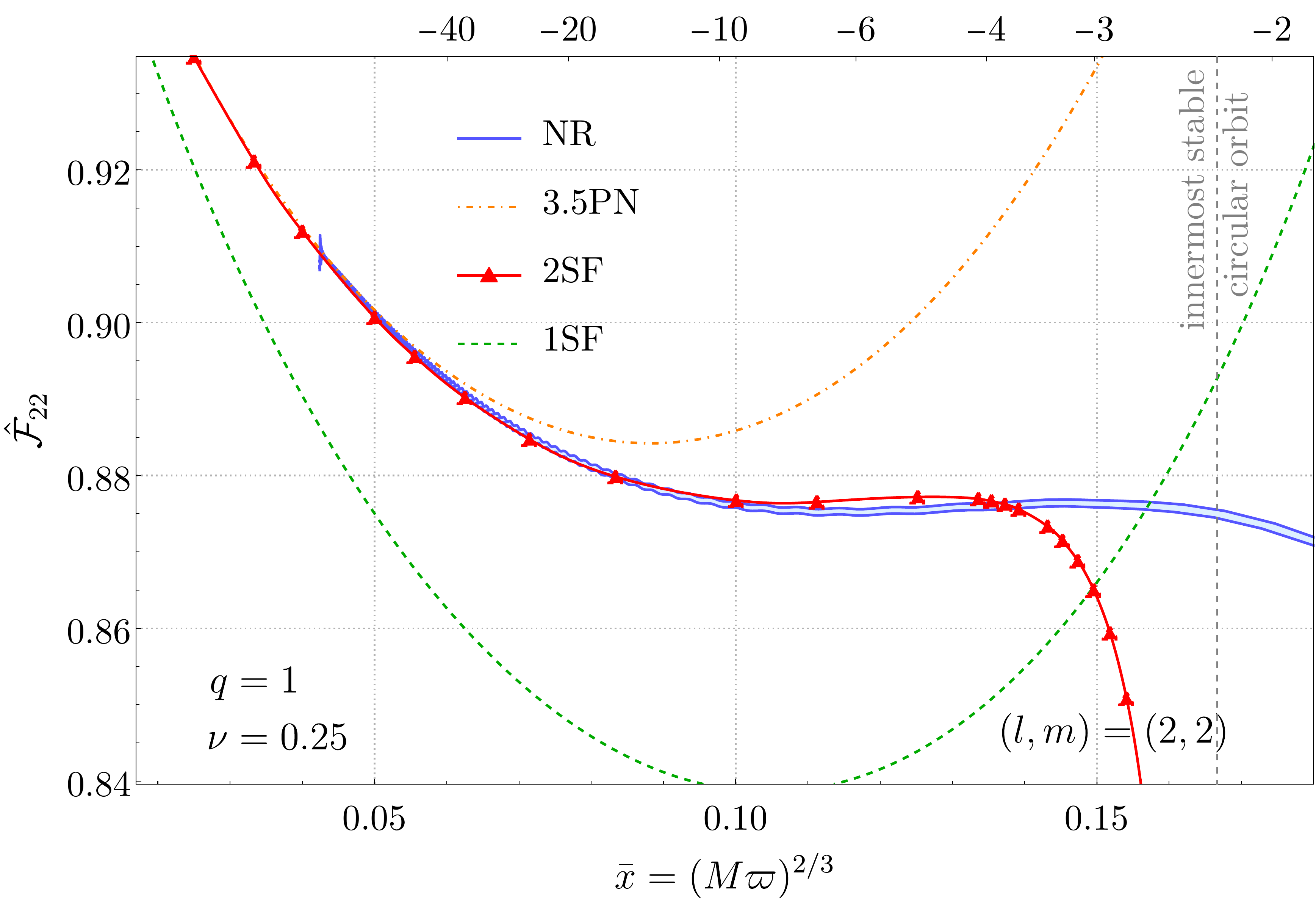}
	\caption{The same as Fig.~\ref{fig:GSFvsNRvsPN_l2m2} but with $q=1$. Despite being a small-$\e$ (small-$\nu$, large-$q$) expansion, the 2SF result agrees remarkably well with the NR flux for this equal-mass binary. The NR flux was computed from SXS:BBH:1132 \cite{Boyle:2019kee}.}\label{fig:GSFvsNRvsPN_l2m2_q1}
\end{figure}

\textit{Flux calculation.}
To facilitate comparisons, we parametrize the NR and GSF fluxes in terms of quantities that can be computed directly from the waveform~\cite{Boyle:2008ge}. 
We decompose the waveforms as $h(t) = h_+ + i h_\times  = r^{-1}\sum_{lm} h_{lm}\, {}_{-2}Y_{lm}(\theta,\phi)$, where ${}_{-2}Y_{lm}$ is a spin-weight $-2$ spherical harmonic.
We further decompose each mode into an amplitude and a phase, $h_{lm}(t) = A_{lm}(t) e^{i\Phi_{lm}(t)}$, where $A_{lm}(t)$ and $\Phi_{lm}(t)$ are real functions. 
The flux is $\F_{lm}(t) = \frac{1}{16\pi}|\dot{h}_{lm}(t)|^2$ and the frequency is defined as $\varpi = \dot{\Phi}_{2,2}/2$, where an overdot denotes $d/dt$.
In the weak field $\varpi \simeq \Omega$, and for small mass ratios $\varpi=\Omega+\mathcal{O}(\e)$.
We then define inverse orbital separations $x(t) = (M\Omega)^{2/3}$ and $\bar{x}(t) = (M\varpi)^{2/3}$.
It will be useful to define the Newtonian-normalized flux $\nF_{lm} \equiv \F_{lm}/\F_{lm}^\text{N}$, where $\F_{lm}^\text{N}$ is the leading term in the PN series for that mode; e.g., $\F_{22}^\text{N} = 32x^5\nu^2/5$, $\F_{33}^\text{N} = 243x^6\nu^2(1-\nu^2)$, etc.~\cite{Messina:2018ghh}.

In our GSF calculation, $\F_{lm}$ is calculated from the $lm$ mode of $(\e h^{1,m}_{\alpha\beta}+\e^2 h^{2,m}_{\alpha\beta})$ at null infinity~\cite{Barack-Lousto:05,Akcay:11}; since $\mathcal{F}_{lm}$ only depends on $d\phi_p/dt=\Omega$, we can calculate $\mathcal{F}_{lm}(\e,\Omega)$ without knowing $\phi_p(t)$.
We write it as $\F_{lm}^{\text{SF},\e}(\epsilon,\Omega) = \e^2 \F_{lm}^{\text{SF,}1\e}(\Omega) + \e^3 \F_{lm}^{\text{SF,}2\e}(\Omega) + \mathcal{O}(\e^4)$.
For comparable-mass binaries, it is natural (and it is known to improve BHPT's accuracy) to express GSF results in terms of the symmetric mass ratio, $\nu$~\cite{LeTiec-etal:11,vandeMeent:2020xgc}.
We hence write $(\e,\Omega)$ as functions of $(\nu,x)$ and re-expand our flux as $\F_{lm}^\text{SF}(\nu,x) = \nu^2 \F_{lm}^\text{SF,1}(x) + \nu^3 \F_{lm}^\text{SF,2}(x) + \mathcal{O}(\nu^4)$.
Finally, we convert from $x$ to $\bar{x}$ (though we find this correction to be very small for all $\nu$ and $x$ we have considered).

\begin{figure}
	\includegraphics[width=8.5cm]{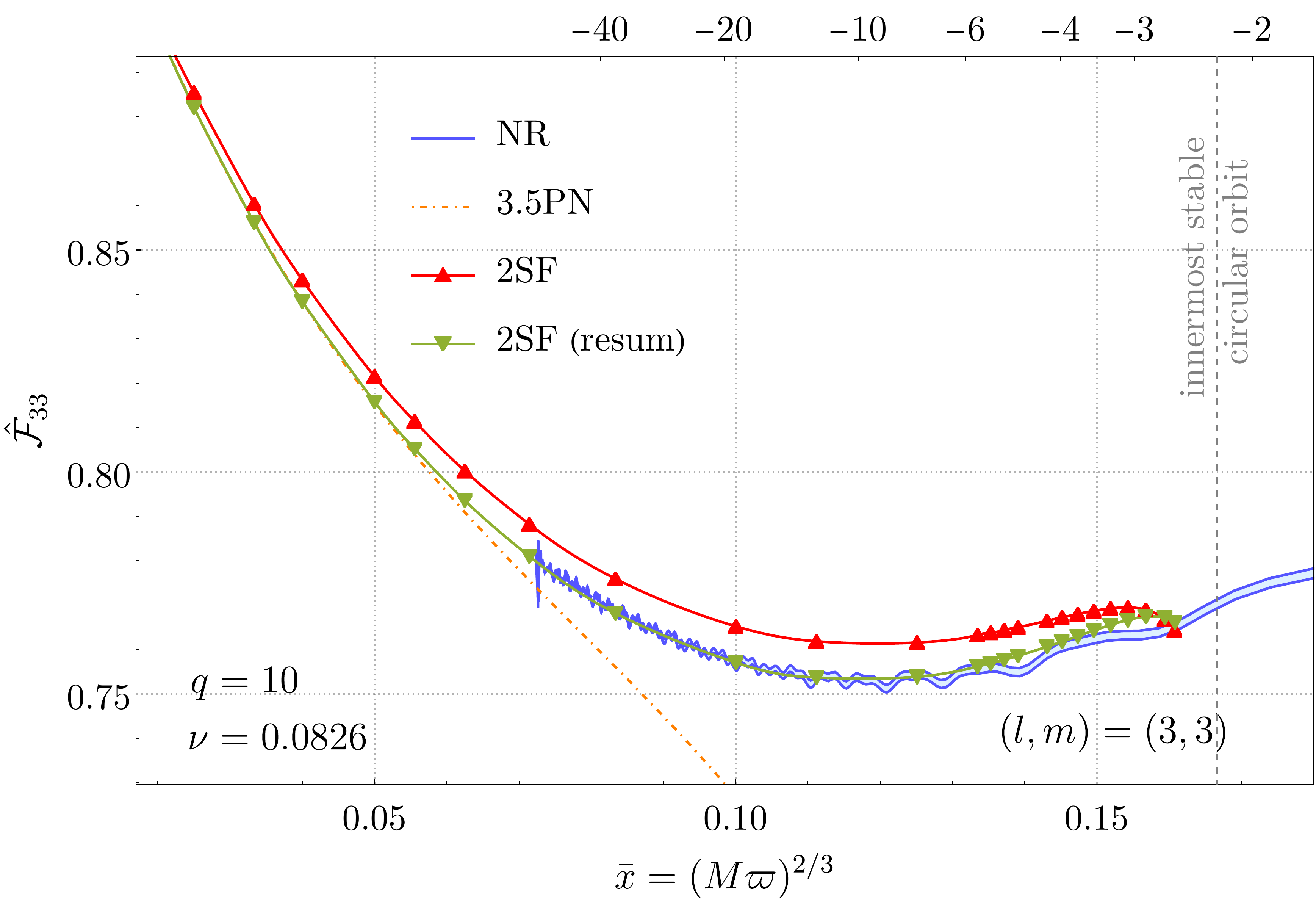}
	\caption{The same as Fig.~\ref{fig:GSFvsNRvsPN_l2m2} but for the $(l,m) = (3,3)$ mode. We see that the 2SF flux does not agree as well with the NR result as in the case of the $(2,2)$-mode. 
	 The simple resummation of the 2SF flux described in the main text results in a substantial improvement in the comparison with the NR flux. 
The relative difference between the NR and resummed 2SF flux up to five cycles before the waveform peak is below $4.5\times10^{-3}$, compared to $1.3\times10^{-2}$ for the nonresummed case. 
The resummation gives similar improvements for the other modes we have computed up to $l=5$. The 1SF result is not visible on the scale of the plot.}\label{fig:Flux_comparison_l3m3_q10}
\end{figure}

\textit{Comparison with numerical relativity simulations for nonspinning binaries.}
With the above definitions we computed the flux from nonspinning NR simulations in the public Simulating eXtreme Spacetimes (SXS) catalogue \cite{Boyle:2019kee}.
The SXS data is provided at different simulation resolutions and the waveform is computed using different extrapolations of finite-radius data to null infinity \cite{Boyle:2009vi}.
We find that the extrapolation order dominates the uncertainty in the NR waveforms, and so in all our comparisons we use the highest resolution NR data and plot the flux computed from the two highest extrapolation orders.
Comparisons between the NR, PN, and GSF fluxes for the $(2,2)$ mode are shown for $q=10$ and $q=1$ in Figs.~\ref{fig:GSFvsNRvsPN_l2m2} and \ref{fig:GSFvsNRvsPN_l2m2_q1}, respectively.
Despite being a small-$\e$ (large-$q$) expansion, we observe that the 2SF flux agrees remarkably well with the NR flux for the dominant $(2,2)$ mode.
For example, for $q=10$ the relative disagreement between the 2SF and NR fluxes  remains below $1.9\times10^{-3}$ until five cycles before the peak amplitude of the waveform.
Even for $q=1$ the relative disagreement  remains below $2.5\times10^{-3}$ until five cycles before the waveform peak.
Closer to the ISCO, the disagreement blows up as a consequence of our two-timescale expansion breaking down.

\begin{figure}
	\includegraphics[width=8.5cm]{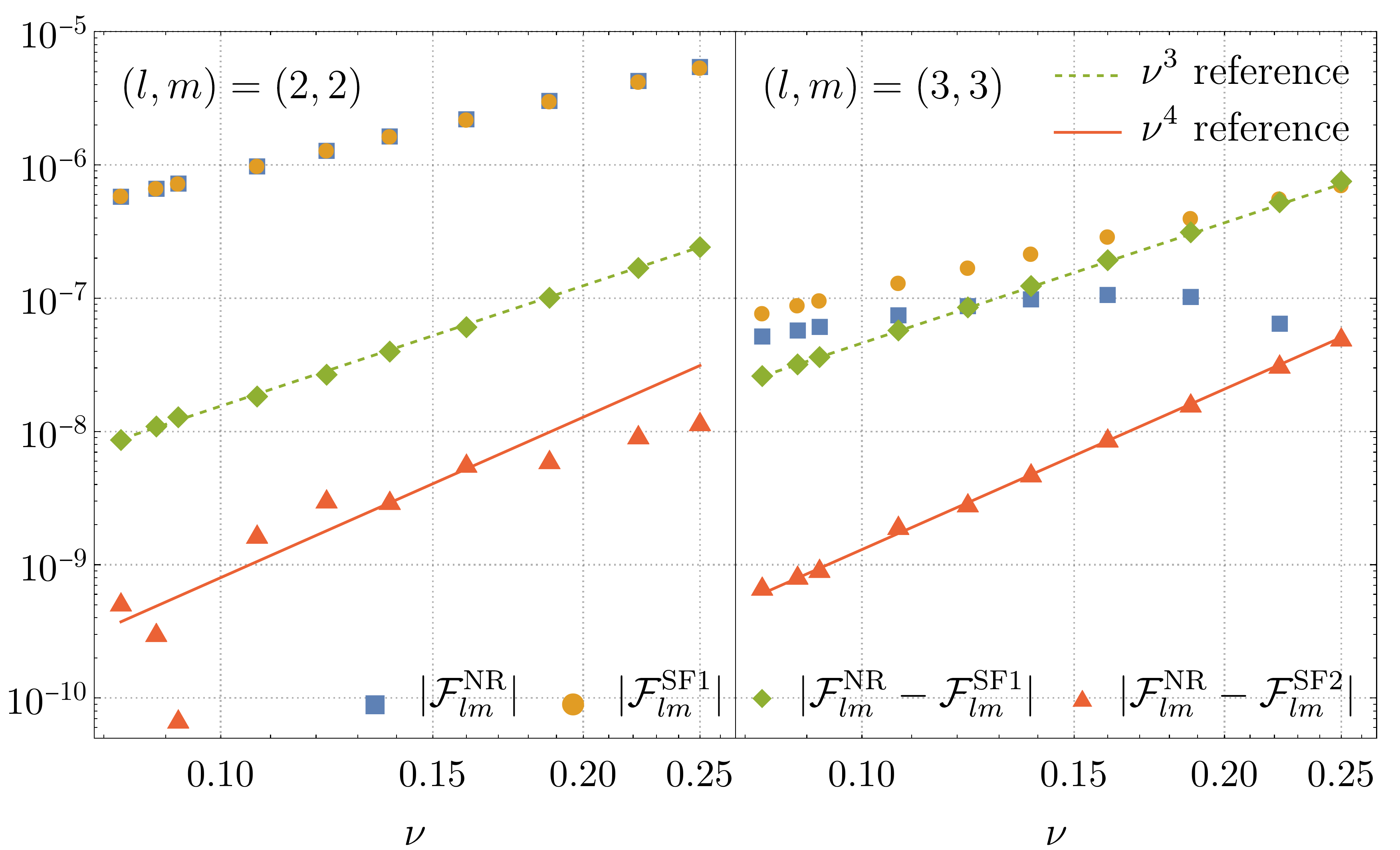}
	\caption{Comparison of NR and GSF fluxes at $\bar{x}=1/9$ for the (2,2)-mode (left panel) and (3,3)-mode (right panel). At leading order both the NR flux (blue squares) and GSF flux (orange circles) scale as $\nu^2$. After subtracting the 1SF flux from the NR flux we find that the residual follows the dashed (green) $\nu^3$ curve. After further subtracting the 2SF fluxes we expect the residual to scale as $\nu^4$ (shown as a solid red curve). For the (2,2)-mode the residual broadly follows the $\nu^4$ trend, but the comparison is complicated by small oscillations in the NR waveform (likely from residual eccentricity and/or centre-of-mass motion in the NR simulation~\cite{SXS:private,Mitman:2021xkq}). For the (3,3)-mode the residual is less subdominant and clearly follows the expected $\nu^4$ behavior. The SXS datasets used in this comparison are listed in the Supplemental Material.}\label{fig:NR_minus_SF}
\end{figure}

For subdominant modes the agreement between NR and 2SF worsens -- see Fig.~\ref{fig:Flux_comparison_l3m3_q10}.
This is not unexpected; by examining the PN series (given in Appendix A of Ref.~\cite{Messina:2018ghh}) we see that for the $(2,2)$-mode the third-order, $\mathcal{O}(\nu^4)$, corrections appear at (relative) 2PN order whereas for the $(3,3)$-mode the first $\mathcal{O}(\nu^4)$ term appears at (relative) 1PN order.
For other modes $\mathcal{O}(\nu^4)$ terms can appear in the leading PN term.
Interestingly we find that the following simple resummation provides a substantial improvement in the accuracy of the GSF flux: $\F_{lm}^\text{SF,resum}(x) = \left[\F_{lm}^\text{SF}(x)/\F_{lm}^\text{N}(x) + \mathcal{O}(\nu^2)\right]\F_{lm}^\text{N}(x)$, where the fraction in brackets is re-expanded through order $\nu$.
This resummed series is constructed to have the property that $\lim_{x\rightarrow0} \hat{\F}_{lm}^\text{SF,resum}(x) = 1$.
Figure~\ref{fig:Flux_comparison_l3m3_q10} shows that it works remarkably well.
Similar results are observed for smaller values of $q$ and/or more subdominant modes.

Furthermore, despite the weaker agreement between the NR and nonresummed 2SF results for the subdominant modes, the total flux (summed up to $l=5$) still compares very well between the two methods as the $(2,2)$-mode dominates the sum. For example, we find that for $q=10$ the relative difference in the total flux  remains below $3.2\times10^{-3}$ up until five cycles before the waveform amplitude peak.

Finally, we compare the GSF and NR results as a function of $\nu$ in Fig.~\ref{fig:NR_minus_SF}. 
Both the 1SF and NR flux scale as $\mathcal{O}(\nu^2)$, and after subtracting the 1SF from the NR flux we observe that the residual falls off as $\nu^3$.
After further subtracting the 2SF flux we find that the residual scales as $\nu^4$.
This gives us confidence that our GSF result captures the behaviour of the full NR flux through $\mathcal{O}(\nu^3)$.
Our results also suggest that by comparing 2SF and NR fluxes, it may be possible to numerically extract the third-order, $\mathcal{O}(\nu^4)$ flux.

\textit{Flux from spinning binaries.}
Our expansion in Eq.~\eqref{eq:pert} allows us to include a small, $\mathcal{O}(\e)$ spin on the primary, which evolves due to absorption of GWs during the inspiral but can take any (small) initial value. Furthermore, we can also consistently add corrections due to a spinning secondary so long as its angular momentum per unit mass is of $\mathcal{O}(\e)$, as is the case for a compact secondary.

\begin{figure}
	\includegraphics[width=8.5cm]{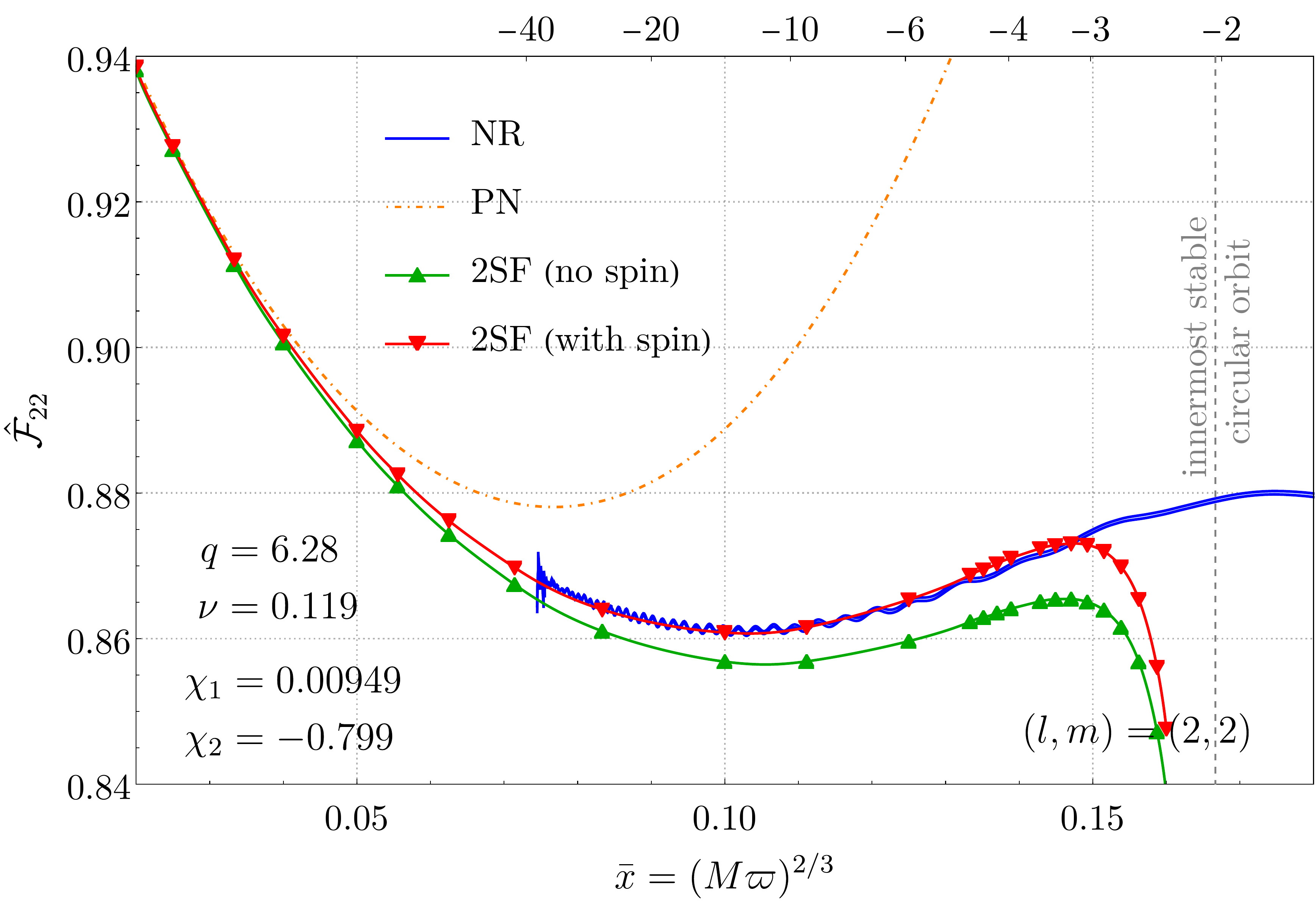}
	\caption{Flux comparison for a spinning secondary with $q\simeq6.28$. The primary has a very low spin and the secondary has a retrograde spin with $\chi_2 \simeq -0.8$. The NR flux is computed from SXS:BBH:1436 \cite{Boyle:2019kee}. The PN flux \cite{Messina:2018ghh} is shown with the dot-dashed (orange) curve. The 2SF flux without spinning flux corrections is shown by the (green) triangles. After the spining flux corrections are added the 2SF result (red, upside-down triangles) agrees well with the NR flux.}\label{fig:flux_spinning_secondary}
\end{figure}

To facilitate comparisons between GSF and NR fluxes with spins, we follow Ref.~\cite{Messina:2018ghh} and introduce $X_1=(1+\sqrt{1-4\nu})/2$ and $X_2 = 1 - X_1$. 
With these, we define $\tilde{a}_i = a_i/M = X_i \chi_i$ and $\F_{lm}^{\text{SF,spin}}(\bar x)  = \F^\text{SF}_{lm}(\bar x) + \sum_{i=1}^2 \tilde{a}_i \F^{\text{spin,}i}_{lm}(\bar x)$, where $\F^{\text{spin,}i}_{lm}(\bar x)$ is the leading contribution to the flux due to the spins (discussed below).

We first consider binaries with a spinning secondary and a nonspinning primary. 
In perturbation theory, many authors have computed $\F^{\text{SF,spin,}2}_{lm}(\bar x)$ for circular orbits \cite{Nagar:2019wrt,Akcay:2019bvk,Piovano:2020zin}. 
Here we make use of the results of Ref.~\cite{Akcay:2019bvk}, where the linear-in-spin flux is computed as a function of the orbital frequency.
As before, even for a small-$q$ binary and a rapidly rotating secondary, we find good agreement with NR simulations -- see Fig.~\ref{fig:flux_spinning_secondary}.

\begin{figure}
	\includegraphics[width=8.5cm]{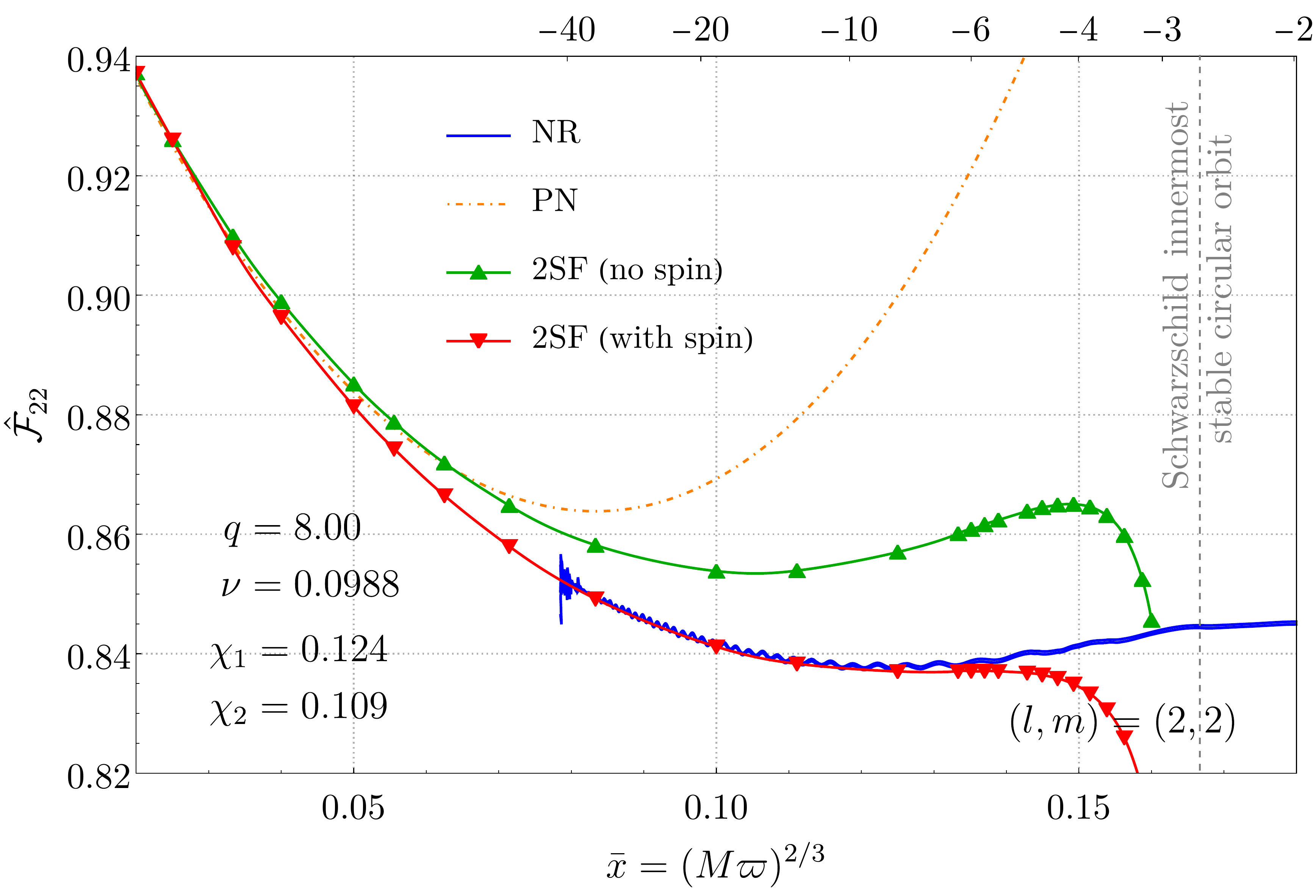}
	\caption{The same as Fig.~\ref{fig:flux_spinning_secondary} but with a slowly spinning primary with $q\simeq 8$. The primary is spinning with $\chi_1 \simeq 0.12$ and the secondary has $\chi_2 \simeq 0.11$. The NR flux is computed from SXS:BBH:1460 \cite{Boyle:2019kee}.}\label{fig:flux_spinning_primary}
\end{figure}

We next consider a spinning primary with  $\tilde{a}_1 \sim \e$, putting the spin in the perturbation  $h^{1,0}_{\alpha\beta}$ in Eq.~\eqref{eq:pert}. We find that the resulting correction to the flux agrees with the linear-in-$a_1$ flux extracted from first-order calculations on a Kerr background \cite{Hughes:1999bq} to within $4.2\times10^{-5}$ (relative). 
If we add this contribution to the 2SF flux we again find good agreement with NR when the primary is slowly rotating -- see Fig.~\ref{fig:flux_spinning_primary}.

\begin{figure}
	\includegraphics[width=8.5cm]{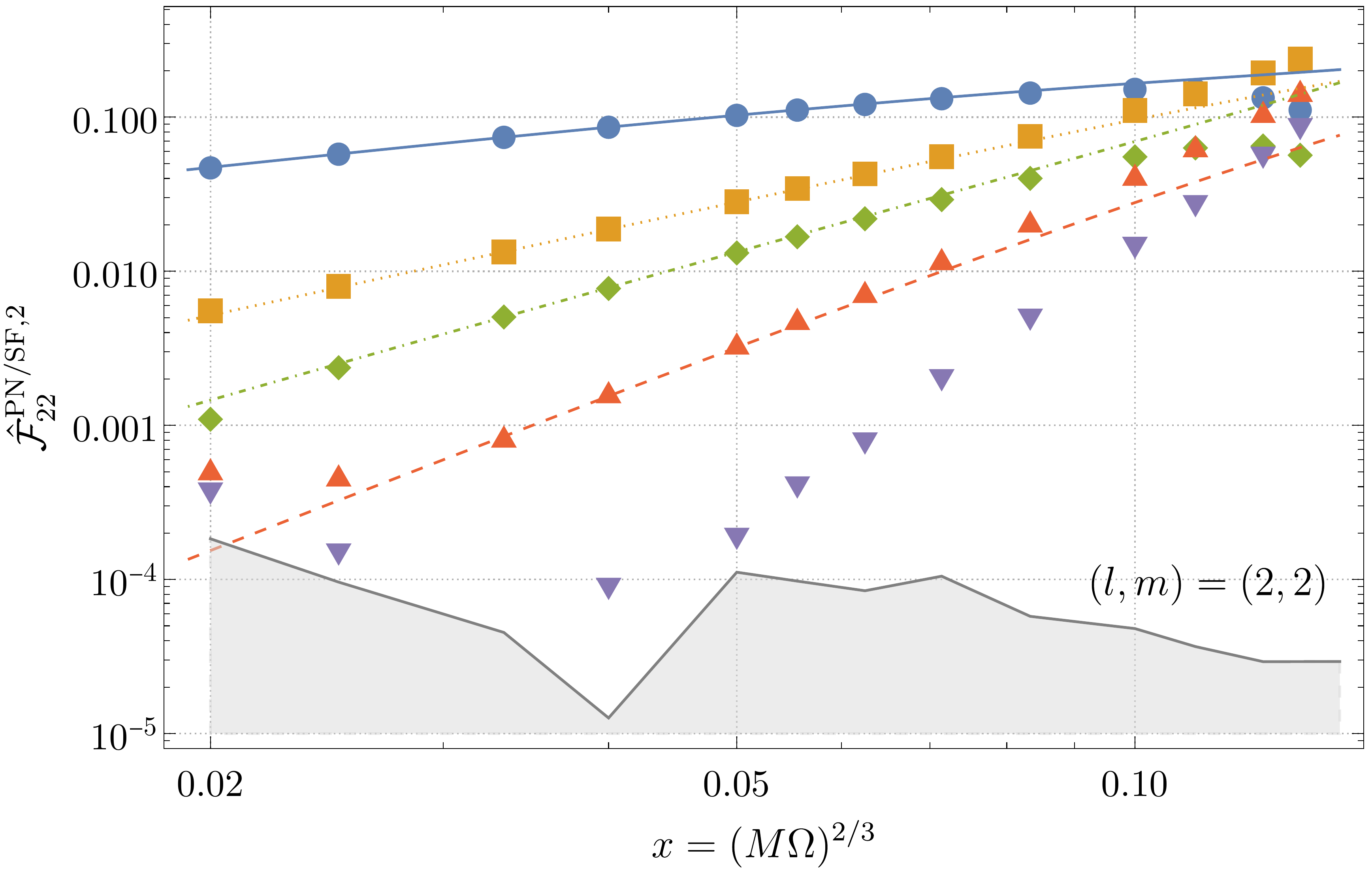}
	\caption{The (Newtonian-normalized) second-order flux, $\hat{\F}^{\text{SF,}2}$, vs the $\mathcal{O}(\nu^3)$ contributions to the PN series, $\hat{\F}^{\text{PN,}2}$, for the $(l,m)=(2,2)$ mode.
	The solid (blue) curve shows the $\mathcal{O}(\nu^3)$ contribution to the 3.5PN flux and the circles show our 2SF result. We then subtract the leading $\mathcal{O}(\nu^3)$ PN term, $55x/21$, from both and get the (orange) dotted curve and the squares. After subtracting all the PN terms through $x^2$, we get the (green) dot-dashed and diamonds. Subtracting all the PN terms through $x^3$, we get the (red) dashed curve and the triangles. Finally, after subtracting all the PN terms through $x^{7/2}$, i.e., all the known PN terms, the subdominant residual is shown with (purple) upside down triangles. The residual appears to approach the $x^4$ reference curve shown by the long dashed (purple) line.  The gray shaded region shows the estimated error in our 2SF flux. We see similar agreement for other modes. }\label{fig:flux_vs_pn}
\end{figure}

\textit{Comparison with post-Newtonian theory.}
In the weak field, we can cleanly compare our GSF flux results against analytic PN expansions that can be determined from the GW amplitudes \cite{Blanchet:2008je,Faye:2012we, Faye:2014fra, Messina:2018ghh}.
Comparing the $\mathcal{O}(\nu^3)$ terms in the PN series to $\F^\text{SF,2}_{lm}(x)$, we find agreement with all known terms through 3.5PN -- see Fig.~\ref{fig:flux_vs_pn}.

\textit{Conclusions.}
For the first time, we have computed the gravitational-wave energy flux to future null infinity for compact binaries in quasicircular orbits, through second order in the  binary's mass ratio.
We find that the results agree remarkably well with fluxes computed for comparable-mass binaries via numerical relativity.
It is well known that second-order results are crucial for EMRI science \cite{Hinderer-Flanagan:08}, and our results strongly suggest self-force calculations will be effective in modelling IMRIs.

There are many directions in which the present work can be extended.
The most immediate is the computation of the local self-force, with which we can evolve the orbital phase (which did not enter into the flux) and compute the associated waveform.
The orbital phase might also be computed using the flux presented here combined with an appropriate energy-balance law, as in PN waveform templates~\cite{Isoyama:2020lls}.
Once the metric perturbation~\eqref{eq:pert} is computed for nonoscillatory ($m=0$) modes we can also construct second-order conservative corrections to the dynamics~\cite{Pound:14c}, providing gauge-invariant inputs for other approaches to the relativistic two-body problem \cite{Bini-Damour:16,Bini:2019nra}.

Our two-timescale expansion breaks down near the ISCO.
This can be overcome  by matching the two-timescale expansion to a transition to plunge \cite{Ori-Thorne:00, Apte:2019txp,Compere:2021iwh}.
Further attaching a post-merger approximation based on a quasi-normal mode expansion will then provide complete inspiral-merger-ringdown waveforms.

Astrophysically, we expect many supermassive black holes in EMRI binaries to be rapidly spinning \cite{Babak-etal:17}.
Unfortunately, our present calculation does not easily extend to Kerr spacetime as the equations for the Lorenz-gauge metric perturbation have no known separable form.
Multiple parallel efforts are underway to address this \cite{spiers_etal,Toomani:2021jlo,Dolan:2021ijg}.
EMRIs are also expected to have considerable eccentricity near merger \cite{Hopman:2005vr}, and work is underway to develop 2SF techniques for these binaries~\cite{Leather_in_prep}.


\begin{acknowledgments}
\textit{Acknowledgements.} We thank Alexandre Le Tiec and Alessandro Nagar for helpful conversations. AP acknowledges support from a Royal Society University Research Fellowship, a Royal Society Research Fellows Enhancement Award, and a Royal Society Research Grant for Research Fellows. NW gratefully acknowledges support from a Royal Society - Science Foundation Ireland University Research Fellowship. This material is based upon work supported by the National Science Foundation under Grant Number 1417132. This work makes use of the Black Hole Perturbation Toolkit \cite{BHPToolkit} and Simulation Tools \cite{SimulationToolsWeb}.
\end{acknowledgments}

\bibliography{bibfile}

\end{document}


\title{Supplemental material} 

\maketitle

\setcounter{equation}{0}
\setcounter{figure}{0}
\setcounter{table}{0}
\makeatletter
\renewcommand{\theequation}{S\arabic{equation}}
\renewcommand{\thefigure}{S\arabic{figure}}
\renewcommand{\bibnumfmt}[1]{[S#1]}
\renewcommand{\citenumfont}[1]{S#1}

In Fig.~4 where we compare the NR and GSF fluxes as a function of the symmetric mass ratio we use the following datasets from the SXS catalogue \cite{Boyle:2019kee}:

\begin{table}[!htb]
\begin{tabular}{l | l l | l l }
dataset name & $q$  & $\nu$ & $\chi_1$ & $\chi_2$\\
\hline
SXS:BBH:1132 & 1.000 	&   0.2500 	& $-1.14\times10^{-7}$			& $-1.14\times10^{-7}$			\\
SXS:BBH:1165 & 2.000 	&   0.2222 	& $-7.91\times10^{-5}$			& $\;\;\;1.95\times10^{-5}$		\\
SXS:BBH:2265 & 3.000 	&   0.1875 	& $\;\;\;2.24\times10^{-6}$		& $\;\;\;5.41\times10^{-6}$		\\
SXS:BBH:1220 & 4.001 	&	0.1600 	& $-5.63\times10^{-5}$			& $\;\;\;3.31\times10^{-5}$		\\
SXS:BBH:0187 & 5.039	&	0.1382	& $\;\;\;8.80\times10^{-6}$		& $-1.20\times10^{-5}$			\\
SXS:BBH:0181 & 6.000 	&	0.1225	& $-4.34\times10^{-6}$			& $-9.28\times10^{-6}$			\\
SXS:BBH:0188 & 7.187	&	0.1072	& $\;\;\;1.55\times10^{-6}$		& $-2.44\times10^{-5}$			\\
SXS:BBH:0199 & 8.729	&	0.0922	& $-1.11\times10^{-6}$			& $-3.31\times10^{-5}$			\\
SXS:BBH:1108 & 9.200	&	0.0884	& $-2.25\times10^{-6}$			& $-1.46\times10^{-6}$			\\
SXS:BBH:1107 & 10.000	&	0.0826  & $\;\;\;3.65\times10^{-6}$		& $\;\;\;5.78\times10^{-8}$
\end{tabular}
\end{table}
These datasets were chosen because (i) they span the range $q=1$ to $q=10$, (ii) the components of the binary are initially very slowly spinning, and (iii) the magnitude of oscillations in the flux are small.
The small oscillations in the flux are likely due to residual eccentricity and/or the motion of the centre of mass of the binary \cite{SXS:private}.

\bibliographystyle{apsrev4-2}
\bibliography{../bibfile}